\pdfoutput=1
\documentclass[conference]{IEEEtran}
\usepackage[comma, square, numbers, sort&compress]{natbib}
\usepackage{amsmath,amssymb,amsfonts}
\usepackage[multiple,flushmargin]{footmisc}
\usepackage{graphicx}
\usepackage{textcomp}
\usepackage{xcolor}
\usepackage{xparse}
\usepackage{xfrac}
\usepackage{relsize}
\usepackage[noend]{algpseudocode}

\usepackage[utf8]{inputenc}
\usepackage{algorithm}
\usepackage[noend]{algpseudocode}
\usepackage{tcolorbox}
\usepackage{color}
\usepackage{xspace}
\usepackage{epsfig,endnotes,color,paralist,multirow}
\usepackage{xspace}
\usepackage{wrapfig}
\usepackage{epsfig}
\usepackage{amsmath}
\usepackage{amsfonts}
\usepackage{graphicx}
\usepackage[normalem]{ulem}
\usepackage{subcaption}
\usepackage{url}
\usepackage{soul}

\usepackage{multirow}
\usepackage{array}
\newcolumntype{H}{>{\setbox0=\hbox\bgroup}c<{\egroup}@{}}

\def\backGroundColor{white}
\def\txtsize{\normalsize}
\def\bl{\par\textcolor{\backGroundColor}{\txtsize{ empty line }}\par}

\def\includeComments{include}
\def\includ{include}
\def\comm[#1]{\ifx\includeComments\includ  \bl \texttt{\textbf{\textit{Note: #1}}} \par \fi}
\def\inlinecomm[#1]{\ifx\includeComments\includ  \textit{Note: #1} \fi}

\def\blackbox{\hfill {\vrule height6pt width6pt depth0pt}}
\def\Box{\hfill \framebox(5.25,5.25){}}

\newcommand{\blue}[1]{\textcolor{blue}{#1}}

\newcommand{\eat}[1]{}
\newcommand{\vsp}{\vspace*{0.1in}}

\newcommand{\para}[1]{\smallskip \noindent {\bf #1}}
\newcommand{\softpara}[1]{\smallskip \noindent \underline{#1}}

\newcounter{theorem}
\newtheorem{thm}{Theorem}

\newtheorem{corollary}{Corollary}

\newtheorem{defin}{Definition}
\newtheorem{ex}{Example}
\newtheorem{lem}{Lemma}
\newtheorem{ob}{Observation}

\newenvironment{theorem}{\begin{thm} \nopagebreak}{{\hfill$\blackbox$} \end{thm}}
\newenvironment{thm-prf}{\vsp \begin{thm} \nopagebreak}{\end{thm}}

\newenvironment{lem-wo-prf-box}{\vsp \begin{lem} \nopagebreak}{\end{lem}}
\newenvironment{lem-prf}{\vsp \begin{lem} \nopagebreak}{\end{lem}}

\newenvironment{cor-prf}{\vsp \begin{corollary} \nopagebreak}{\end{corollary}}

\newcounter{packednmbr}

\newcounter{packedalph}

\newcommand{\po}{\ensuremath{\mathcal{P}}\xspace}

\newcommand{\eps}{\mbox{EP}\xspace}
\newcommand{\epss}{\mbox{EPs}\xspace}
\newcommand{\es}{\mbox{ES}\xspace}

\newcommand{\os}{\mbox{\tt WaitLess}\xspace}
\newcommand{\wt}{\mbox{\tt Waiting}\xspace}
\newcommand{\sls}{\mbox{\tt SLS}\xspace}
\newcommand{\gdsls}{\mbox{\tt GD-SLS}\xspace}
\newcommand{\adsg}{\mbox{\tt APSG}\xspace}
\newcommand{\geng}{\mbox{\tt GG}\xspace}
\newcommand{\clus}{\mbox{\tt CLUS}\xspace}
\newcommand{\naive}{\mbox{\tt Naive}\xspace}
\newcommand{\nosl}{\mbox{\tt Non-SLs}\xspace}
\newcommand{\ack}{\mbox{\tt ACK}\xspace}
\newcommand{\demandT}{\mbox{$\tau$}\xspace}

\DeclareMathOperator*{\argmin}{arg\,min}

\newcommand{\php}{\mbox{$p_{ob}$}\xspace}          % BSM (photon-photon) probability of success.
\newcommand{\bt}{\mbox{$t_{b}$}\xspace}         % BSM (atom-atom time. Inverse is rate.
\newcommand{\bp}{\mbox{$p_{b}$}\xspace}          % BSM (atom-atom probability of success.
\newcommand{\gt}{\mbox{$t_g$}\xspace}      % Node generation (atom-photon) time. Inverse is rate.
\newcommand{\gp}{\mbox{$p_g$}\xspace}       % Node generation probability of success.
\newcommand{\ct}{\mbox{$t_c$}\xspace}      % Classical time.
\newcommand{\entT}[3]{
        \expandafter\ifx\expandafter\relax\detokenize{#1}\relax
            \expandafter\ifx\expandafter\relax\detokenize{#2}\relax
                \ensuremath{\mathrm{T}}\xspace
            \else
                \ensuremath{\mathrm{T_{#2,#3}}}\xspace
            \fi
        \else
            \ensuremath{\mathrm{T^#1_{#2,#3}}}\xspace
        \fi\xspace}
\newcommand{\tauT}[3]{
        \expandafter\ifx\expandafter\relax\detokenize{#1}\relax
            \expandafter\ifx\expandafter\relax\detokenize{#3}\relax
                \ensuremath{\mathrm{\tau}}\xspace
            \else
                \ensuremath{\mathrm{\tau_{#2,#3}}}\xspace
            \fi
        \else
            \ensuremath{\mathrm{\tau^#1_{#2,#3}}}\xspace
        \fi\xspace}

\def\BibTeX{{\rm B\kern-.05em{\sc i\kern-.025em b}\kern-.08em
    T\kern-.1667em\lower.7ex\hbox{E}\kern-.125emX}}
\begin{document}

\title{Pre-Distribution of Entanglements in Quantum Networks}
%\thanks{Supported by NSF, Cisco.}}

\author{\IEEEauthorblockN{Mohammad Ghaderibaneh, Himanshu Gupta, CR Ramakrishnan, Ertai Luo}
\IEEEauthorblockA{\textit{Stony Brook University, Stony Brook, NY, USA}}}

\maketitle

\begin{abstract}
%Quantum Computing has a potential, if realized, to significantly alter the computing landscape. However, building large-scale quantum computers is a key challenge. 
%Quantum Networks (QNs) 
%enable the construction of large distributed quantum computing platforms by connecting smaller quantum
%computers, and enable many novel applications.
%%%%%%%%%%%%%%%%%%%%%%%%%%%%%%%
Quantum network communication is challenging, as the No-Cloning theorem in quantum regime makes many classical techniques inapplicable. 
%%%%%%%%%%%%%%%%%
For long-distance communication, the only viable approach is 
teleportation of quantum states, which requires a prior distribution of 
entangled pairs (\epss) of qubits. 
Establishment of \epss across remote nodes can incur significant 
latency due to the low probability of success of the underlying 
physical processes. 
%%%%
To reduce \eps generation latency, prior works have looked at selection 
of efficient entanglement-routing paths and simultaneous use of multiple such paths for \eps generation.
%%%%%%%%%%%%%%

In this paper, we propose and investigate a complementary technique
to reduce \eps generation latency---to pre-distribute \epss over
certain (pre-determined) pairs of network nodes; these pre-distributed
\epss can then be used to generate \epss for the requested pairs, 
when needed, with lower generation latency. 
%%%%
For such an pre-distribution approach to be most effective, we need
to address an optimization problem of selection of node-pairs where
the \epss should be pre-distributed to minimize the generation latency
of expected \eps requests, under a given cost constraint. In this paper,
we appropriately formulate the above optimization problem and design two
efficient algorithms, one of which is a greedy approach based on an 
approximation algorithm for a special case. 
%%%%%%%%
%%%%%%%%%%
Via extensive evaluations over the NetSquid simulator~\cite{netsquid2020}, we demonstrate the effectiveness of our approach and developed techniques;  
we show that our developed algorithms outperform a naive approach by up to an order
of magnitude. 
\end{abstract}

% \begin{IEEEkeywords}
% Long-distance Entanglement, Generation Latency.
% \end{IEEEkeywords}

\pagenumbering{arabic}
%%%%%%%%%%%%%%%%%%%%%%%%%%%%%%%%%%%%%%%%%%%%%%%%%%%
\pagestyle{plain}

\section{\bf Introduction}
\label{sec:intro}

Fundamental advances in physical sciences and engineering have led to the realization of working quantum computers (QCs)~\cite{google-nature-19, ibm-quantum-roadmap}.  
%Nevertheless, much work remains to fully realize the revolutionary potential of quantum computing. Specifically, 
However, there are significant limitations 
to the capacity of individual QC~\cite{Caleffi+:18}.  Quantum networks (QNs) enable the construction of large, robust, and more capable quantum computing platforms by connecting smaller QCs. 
Quantum networks~\cite{qn}
also enable various important applications~\cite{qsn,qkd,atomic,secure,byzantine}.
%such as quantum sensor networks~\cite{qsn}, quantum key
%distribution (QKD)~\cite{qkd, qkd-2}, 
%atomic clock~\cite{atomic}, distributed consensus~\cite{byzantine}, secure %communications~\cite{secure}, etc. 
%%%%%%%%
However, quantum network communication is challenging---e.g., physical transmission of quantum states across nodes can incur irreparable communication errors, 
as classical procedures such as amplified signals 
or re-transmission cannot be applied due to quantum no-cloning~\cite{wooterszurek-nocloning,Dieks-nocloning}.
At the same time, certain aspects unique to the quantum regime, such as entangled states, enables 
novel mechanisms for communication.
%%%%%%%%%%%%%
In particular, teleportation~\cite{Bennett+:93} transfers quantum states with just classical
communication, but requires an a priori establishment of entangled pairs (\epss).
%%%%
This paper focusses on efficient generation of \epss in a quantum network.

Establishment of \epss over long distances is challenging.
Coordinated entanglement swapping (e.g. DLCZ protocol~\cite{dlcz}) using quantum repeaters 
can be used to establish long-distance entanglements from short-distance entanglements.
However, due to low probability of success of the underlying physical processes
(short-distance entanglements and swappings), \eps generation can incur significant latency---of the order of 10s to 100s of seconds between nodes 100s of kms away~\cite{gisin,caleffi,tens}.
%%%%%%%%%%%%%%%%%%%%%%%
Thus, we need to develop techniques that enable faster generation of long-distance 
\epss. 
%%%%%%%%%%%%%%%%%%%%%%%%%%%%%%%%%%%%%%%%%%%%%%%
In the past, researchers have addressed the above problem of high \eps generation latency
by developing techniques for selection of efficient entanglement routes~\cite{sigcomm20, delft-lp, caleffi} (or swapping trees~\cite{swapping-tqe22}), and proposing to use multiple
such routes/trees for each \eps generation~\cite{sigcomm20, delft-lp, swapping-tqe22}. These approaches are effective, but can lead to heavy and bursty use of network resources at the time of \eps request, and more importantly, the generation latency can still be 
high. 

\para{Pre-Distribution of \epss via Super-Links.} 
In this paper, we propose a complementary {\em pre-distribution} approach to lower the \eps generation latencies: in this pre-distribution approach,
we proactively generate and pre-distribute/store \epss over certain (pre-determined) pairs of network
nodes. When needed, these pre-distributed \epss can then be ``used'' to generate \epss across {\em requested} pair(s) of nodes; use of pre-distributed \epss, if carefully
chosen, can result in lower generation latency for the requested 
\epss~\cite{greedy2019distributed}.\footnote{An extreme pro-active approach could be to continuously generate \epss at the same node pairs as the ones that will be requested (or if unknown, at all the node pairs in the network), but this can be very wasteful of network resources and in many cases, may not be useful or even viable.}  Note that the pre-distribution is complementary and can be used in conjunction with the prior approaches of efficient and multiple entanglement routes. 
%%%%%%%%%%%%%
The pairs of nodes that are chosen for pre-distribution of \epss are referred to as {\em super-links}, as they are tantamount to short-cuts in an entanglement-routing path. 
%%%%%%%%%%%%%%%%%%%%%%%
The above pre-distribution approach is particularly beneficial when we have a priori knowledge about the network traffic (e.g., the distribution of the \epss requests), 
which can be used to select an efficient set of super-links; in general, the
selection of super-links should exploit the ``commonality'' across expected requests.
%%%%%%%%%

\para{Contributions and Organization.}
For the above approach of pre-distribution of \epss to be most effective, we need to 
develop techniques to select an optimal set of super-links. In this context, our paper 
makes the following contributions.

\begin{itemize}
    \item We motivate and formulate the optimization problem of super-link selection (\sls): Given a set of node pairs $\{(s,d)\}$ representing expected \epss requests, 
select a set $S$ of super-links that results in minimum aggregate generation latency 
of the $\{(s,d)\}$ \epss using $S$, under an appropriately defined cost constraint.

\item 
For the above \sls problem, we design two algorithms: (i) Generalized Greedy (\geng) Algorithm, a greedy approach based on an approximation algorithm for a special case
of the \sls problem (\S\ref{sec:greedy}). (ii) Clustering Algorithm, based on the classical $k$-means 
clustering approach (\S\ref{sec:clus}). 

\item
We develop a network protocol (\S\ref{sec:protocol}) for our proposed techniques, and generalize the developed techniques for certain variants of the \sls problem (\S\ref{sec:gen}).

\item 
Using extensive evaluations (\S\ref{sec:eval}), over the NetSquid simulator~\cite{netsquid2020}, we demonstrate
the effectiveness of our approach and developed techniques;  
we show that our algorithms outperform the naive approaches by up to
an order of magnitude.
\end{itemize}
To the best of our knowledge, no prior work has addressed the above optimization problem of selection of super-links for pre-distribution of \epss. The closest work is~\cite{greedy2019distributed} which develops routing algorithms to leverage the pre-distributed entanglements in special network topologies, but does not address the problem of selection of super-links (referred to as {\em virtual links} in~\cite{greedy2019distributed}). We start with presenting the relevant background in the following section (\S\ref{sec:back}).

%%%%%%%%
% In \S\ref{sec:gen}, we discuss various generalizations of the \sls problem and other
% issues. In \S\ref{sec:protocol}, we discuss the network protocol and implementation details. Finally, in \S\ref{sec:eval}, we discuss our empirical results based on 
% extensive evaluations, using the NetSquid simulator~\cite{netsquid2020},
% which demonstrate
% the effectiveness of our approach and developed techniques; we show that our algorithms outperform the naive approaches by a \red{significant margin}. We start with presenting the relevant background in the following section (\S\ref{sec:back}).

\section{\bf Background: Network Model, Entanglement Generation and Latency}
\label{sec:back}

In this section, we present the relevant background related to generation of \epss over remote
nodes in a quantum network. We start with presenting our model of
a quantum network.

\para{Quantum Network.} 
We consider a quantum network (QN) as a graph $G = (V, E)$,
with $V = \{v_1, v_2, \ldots, v_n\}$  and $E = \{(v_i, v_j)\}$
denoting the set of nodes and links respectively. 
A network link is a quantum channel (e.g., using an 
optical fiber or a free-space link),
and, in our context, is used only for establishment 
of link \eps.
%%%%
Pairs of nodes connected by a link are defined as {\em adjacent} nodes. 
%%%%%%%%%%%%%%%%%%%
We follow the network model in~\cite{caleffi,swapping-tqe22} closely.
Fundamentally, each network node $A$ should be capable of generating 
atom-photon \epss, so that it can generate an atom-atom \epss with an 
adjacent node $B$.
%%%%%%%%%%%%%%%%%%%%
% Thus, each node has an atom-photon \eps generator with generation 
% latency (\gt) and probability of success (\gp). Generation latency
% is the time between successive attempts by the node to excite the 
% atom to generate an atom-photon \eps; 
% %this implicitly includes the times for
% %photon transmission, optical-BSM latency, and classical acknowledgement.
{\em For clarity of presentation} and without loss of generality,  
we assume homogeneous network nodes with same parameter values.
% A node's atom-photon generation capacity/rate 
% is its aggregate capacity, and may be split across its incident links 
% (i.e., in generation 
% of \epss over its incident links/nodes).
Each node is also equipped with a certain number of atomic
memories to store the  qubits of the atom-atom \epss. 
%%%%%%%%%%%%%%%%%%
%%%%%%%%%%%%%%%%%
% In particular, a link $e=(A,B)$ is used to transmit telecom-photons 
% from $A$ and $B$
% to the photon-photon BSM device in the middle of $e$.
% Thus, each link is composed of two 
% half-links with a probability of transmission success (\ep) that decreases exponentially with the link distance (see~\S\ref{sec:eval}).
% The optical-BSM operation has 
% a certain probability of success (\php).
To facilitate atom-atom entanglement swapping (\es), each node is also equipped 
with an atomic-BSM device 
with an operation latency (\bt) and probability of success (\bp). Finally, 
there is an independent classical network with a transmission latency (\ct);
we assume classical transmission  
always succeeds.

\para{Quantum Communication via Teleportations and Entanglement Pairs (\epss).}
Direct transmission of quantum data 
is subject to unrecoverable errors, 
as classical procedures such as amplified signals 
or re-transmission cannot be applied due to quantum no-cloning~\cite{wooterszurek-nocloning,Dieks-nocloning}.\footnote{Quantum error correction mechanisms~\cite{muralidharan2016optimal, devitt_2013} can be used to mitigate the transmission errors, 
but their implementation is very challenging and is not expected to be used
until later generations of quantum networks.}
%%%%%%%%%%%%%%%%%%%%%%%%%
An alternative mechanism
for quantum communication is \emph{teleportation}, where a qubit $q$ from a node $A$
is recreated in another node $B$ (\emph{while ``destroying'' the original
qubit $q$}) using only classical communication.
However, this process requires that an \eps 
already established over the nodes $A$ and $B$. 
%%%%%%%%%%%%%%%
Teleportation can thus be used to reliably transfer quantum information.

\para{Generating \epss Across Remote Nodes.}
Establishing an \eps over a pair $(s,d)$ of remote nodes 
consists of two steps: (i) For some path 
$P = (s=x_0, x_1, x_2, \ldots, x_n, d=x_{n+1})$ from $s$ to $d$ in the QN graph
with $x_i$'s being the 
intermediate nodes, first we need to establish \eps between every pair of 
nodes $(x_i, x_{i+1})$ for $0 \leq i \leq n$.
(ii) Then, we conduct a series of entanglement 
swappings over these $(x_i, x_{i+1})$ \epss, to generate an \eps over $(s,d)$.
The sequence in which the entanglement swappings are done has a bearing
on the overall performance; this sequence of entanglement swappings is
best characterized using a swapping tree~\cite{swapping-tqe22} as defined below.

\begin{figure}
    \centering
    \includegraphics[width=0.4\textwidth]{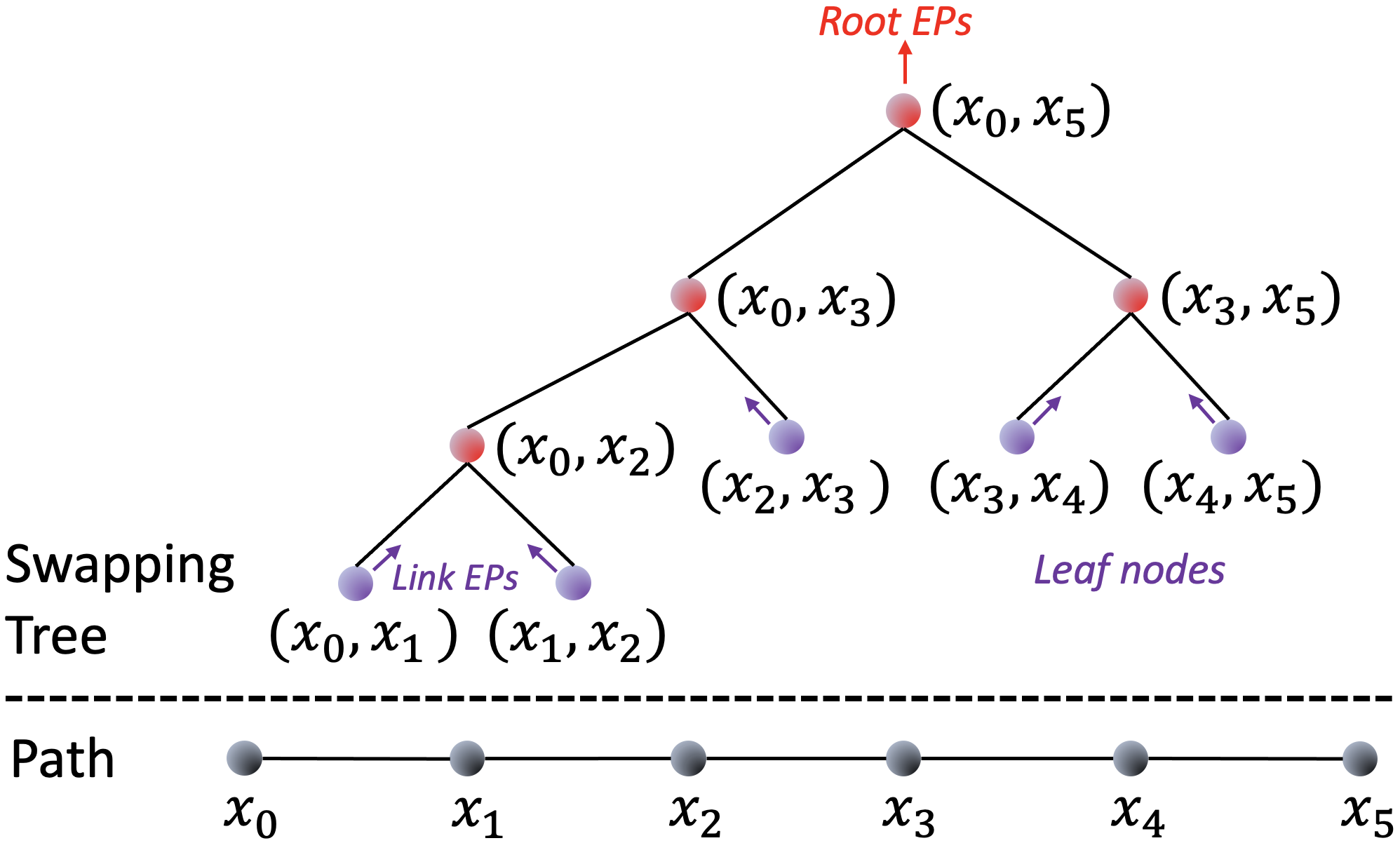}
    %\vspace{-0.1in}
    \caption{A swapping tree over a path. The leaves of the tree are the path-links, which generate link-\epss continuously.}
\vspace{-0.25in}
\label{fig:tree}
\end{figure}

\softpara{Swapping Trees.}
In general, given a path $P = s \leadsto d$ from $s$ to $d$, 
any complete binary tree (called a \textit{swapping tree}) over 
the ordered links in $P$ gives a way to generate an \eps over $(s, d)$.
%%%%%%%%%%%%
Each vertex in the tree corresponds to a pair of network nodes in $P$, 
with each leaf representing a link in $P$. 
%%%%%%%%%%%%%%
Every pair of siblings $(A, B)$ and $(B, C)$ perform an \es over 
$(A,B,C)$  to yield an \eps over $(A,C)$---their parent.
See Fig.~\ref{fig:tree}. 
\emph{Note that subtrees of a swapping tree execute in parallel.}
%%%%%%%%%%%%%%%%%%%%%%%%%%
Different swapping trees over the same path $P$ can have 
different performance characteristics.

\para{Estimating \eps Generation Latency of a Swapping Tree.} 
Since the entanglement swappings are stochastic, we may 
need to repeat the \eps generation process---till it 
succeeds. Overall \eps generation 
latency can be estimated as follows.
%%%%%%%%%%%%%%%%%%%%%%%%%%%
Consider a node $(A,C)$ in the tree, with $(A,B)$ and $(B,C)$ as its two children. 
Let $T_{AB}, T_{BC}$, and $T_{AC}$ be the corresponding (expected) generation latencies. 
%%%%%%%%%%
If  $T_{AB}$ and  $T_{BC}$ are exponentially distributed 
with the same generation latency of $T$, 
then we can derive~\cite{swapping-tqe22} an expression for $T_{AC}$ in terms of $T$.
\begin{equation}
T_{AC} = (\frac{3}{2} T + \bt + \ct)/\bp, 
\label{eqn:tree-rate}
\end{equation}
Above, \bt  and \bp are the operation latency  and probability of success, respectively,
of the 
atomic-BSM device at $B$, and \ct is the transmission latency of the classical bits.
%%%%%%%%%%%%%%%%%
The above makes an assumption of $T_{AB} = 
T_{BC}$, which holds in ``throttled'' trees that uses minimal
underlying network resources~\cite{swapping-tqe22}.
%%%%
To complete our estimation of generation latency, we also need to estimate
\eps generation latencies at the leaves of the swapping tree (i.e., network links).
Link \eps generation latency can be estimated using node and link parameters, 
as derived in~\cite{swapping-tqe22}; we omit the tedious details, as they are not important or
very relevant to our work.
%%%%%%%%
Finally, we note that above we have implicitly assumed a \wt protocol of generation
\epss wherein a qubit of an \eps waits (being stored in a quantum memory) for its counterpart to become available so that an \es operation can be performed. In the alternate protocol, viz., \os, all the links \epss are
synchronized and BSMs conducted simultaneously---obviating the need for memories with higher decoherence times.
We assume the \wt protocol, as it is a more efficient approach for generating remote \epss. 
However, our techniques are also applicable to the \os model.

\section{\bf Super-Links Selection (\sls) Problem}
\label{sec:sls}

In this section, We motivate and formulate the problem of selection of super-links addressed in our paper. 

\para{Pre-distribution of \epss.}
Consider a set of source-destination pairs $\{(s,d)\}$ that require an \eps shared across them, 
from time to time. The \eps across an $(s,d)$ pair may be required to teleport a qubit state from 
$s$ to $d$, as an example. Without any pre-existing \epss, distributing an \eps over $(s, d)$ can
incur very high latency due to stochastic underlying processes with low-probability success; in fact,
the latencies can be of the order of 10s of seconds~\cite{caleffi,tens}. 
%%%%%%%%%
One approach to reduce the \eps generation latency is to use multiple independent swapping-trees simultaneously---this approach has been explored in recent works~\cite{sigcomm20,swapping-tqe22,delft-lp}.
%%%%%%%%%
In this paper, we propose a rather {\em proactive} approach of latency reduction,
wherein we proactively generate and distribute \epss at appropriately 
selected pairs of nodes (not necessarily, the original source-destination pairs).\footnote{This proactive approach of pre-distributing \epss is complementary to the multiple paths/trees approach,
and can be used together with it (see \S\ref{sec:conc}).}
%%%%%%%%%%%%%%%%
Such pre-distribution of \epss is done in a way so that the latency incurred in a real-time
\eps request is minimized. In the extreme case, we could 
pre-distribute \epss directly over the expected $(s,d)$ pairs, which would result in zero 
latency of teleportation requests, but such a strategy would be wasteful
of network resources. Thus, in this paper, we address an optimization problem
of determining an efficient pre-distribution strategy. 
We start with a definition below.

\para{Super-Links (SL).}
Consider a quantum network, and a pair of connected nodes $A$ and $B$. 
Let us decide to continuously generate and maintain \epss over $(A,B)$,
to aid in future \eps requests across certain network node pairs. 
%%%%%%%%%%%%%%%%%
We call $(A,B)$ a {\em super-link (SL)}, since the pair $(A, B)$ can 
essentially be now looked upon as a ``link'' that is generating \epss 
continuously generated over it.\footnote{Prior work~\cite{greedy2019distributed} has
used the term virtual links, instead.} 
{\bf Each super-link $(A,B)$ is implicitly associated
with a path $P_{AB}$} and a swapping-tree over $P_{AB}$ which is used to generate
the \epss over $(A,B)$; in this work, we determine the optimal path and swapping-tree over it using the schemes in our recent work~\cite{swapping-tqe22}.
%%%%%%%%%%%%%%%%%%%%
The main purpose of super-links is to help reduce
the latency of real-time \eps requests over other network nodes. 
%%%%%%%%%%%%%%%%%%%%

\begin{figure}
    \centering
    \includegraphics[width=0.5\textwidth]{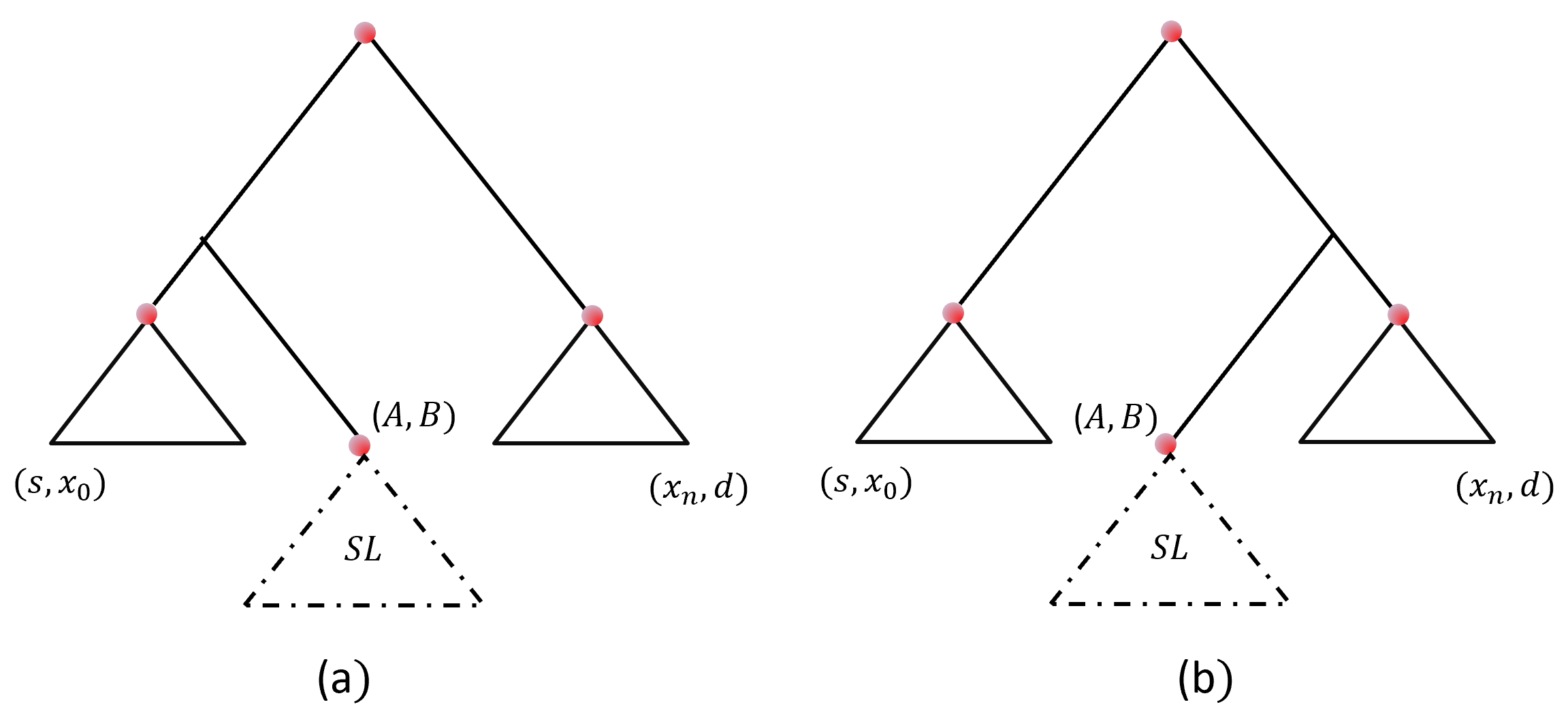}
    % \vspace*{2in}
    \caption{Two possible swapping trees to generate \eps over $(s,d)$ pair using a super-link $(A,B)$.}
    \label{fig:sl-tree}
\end{figure}

To illustrate the usefulness of an SL, 
consider a source-destination pair $(s,d)$ of network nodes
and let $T$ be the expected latency incurred in generating an \eps over $(s,d)$
without any SL.
%%%%%%%
Consider the super-link $(A,B)$, and assume that there are multiple \epss available 
over $(A,B)$ at any point of time. 
%%%%%%%%%%%%%
Then, an \eps over $(s,d)$ can possibly be generated with latency less than $T$ by 
using the \epss over $(A,B)$, depending on the paths $(s, A)$, and $(d,B)$.
In particular, we can first establish \epss over $(s,A)$ and $(B,d)$ using appropriate
swapping trees, and then use a separate swapping tree to generate an $(s,d)$ \eps using \epss from $(s,A)$, 
$(A,B)$, and $(B,d)$. See Fig.~\ref{fig:sl-tree}.
Below, we derive the latency incurred in such an SL-based strategy.
%%%%%%%%%%%%
Note that this overall ``path'' $(s, A, B, d)$ may be very 
different than the shortest/optimal path between $s$ and $d$ to generate an $(s,d)$ \eps.

\para{\eps Generation Latency using a SL.} 
We can use a modified Eqn.~\eqref{eqn:tree-rate} to compute the expected latency
of the swapping trees shown in Fig.~\ref{fig:sl-tree} which use \epss already available
at SL $(A,B)$ to generate an \eps at $(s,d)$. 
%%%%%%%%%%%%%%%%%%%%%%%%
If $T_{s,A}$ is the latency to generate a successful
\eps over $(s,A)$ (using some swapping-tree) 
and $\bp$ is the atomic-BSM probability of success, 
then the expected latency
to generate a successful \eps over $(s,B)$ is $T_{s,B} = (T_{s,A}+ \bt + \ct)/\bp$ if we assume 
sufficiently many \epss
available over $(A,B)$. 
Then, the expected latency to generate an \eps over $(s,d)$ using the
swapping tree in Fig.~\ref{fig:sl-tree}(a) is 
\begin{equation}
T_{sd} = (\frac{3}{2} \max((T_{s,A} + \bt + \ct)/\bp, T_{B,d}) + \bt + \ct)/\bp
\label{eqn:sl-latency-1}
\end{equation}
based on a slight modification of Eqn.~\eqref{eqn:tree-rate}.
Similarly, the expected latency to generate an \eps over $(s,d)$ using the
swapping tree depicted in Fig.~\ref{fig:sl-tree}(b) can be given by:
\begin{equation}
T_{sd} = (\frac{3}{2} \max(T_{s,A}, (T_{B,d} + \bt + \ct)/\bp) + \bt + \ct)/\bp
\label{eqn:sl-latency-2}
\end{equation}
The overall effective expected latency to generate an $(s,d)$ \eps is the minimum of the
above two quantities, as the better of the two swapping-trees in Fig.~\ref{fig:sl-tree} 
would be chosen.

\softpara{Number of \epss Required at Each SL.} 
Above we assumed that sufficiently many \epss are available at an SL. However, 
it is easy to see that the expected
number of \epss used to generate an $(s,d)$ \eps is $1/(\bp)^2$ --- this holds for either
of the swapping trees in figure. This is true since the expected number failures that an 
$(A,B)$ \eps may be involved in is $\bp^2$ as the height of either tree is 2. 
%If there are multiple SLs in a path, the number of \eps required is higher -- see section 7.

\begin{figure}
\vspace*{-0.2in}
    \centering
    \includegraphics[width=0.45\textwidth]{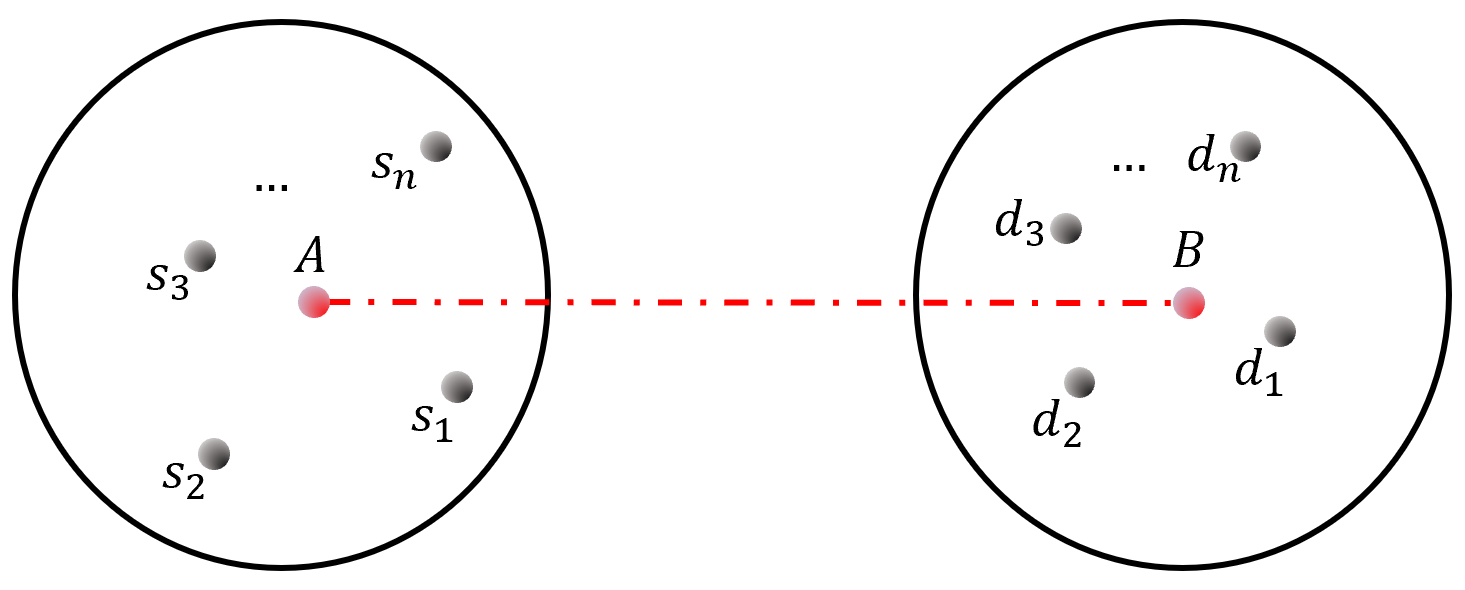}
    % \vspace*{2in}
    \caption{A single super-link $(A,B)$ can be very effective in handling {\em all} \eps requests of the type $(s_i, d_j)$.}
    \vspace*{-0.1in}
    \vspace*{-0.2in}
    \label{fig:sl-cluster}
\end{figure}

\para{Motivation for Selection of SLs.}
The above discussion illustrates how a super-link can be useful in reducing latency.
However, an SL-based \eps generation scheme can also be wasteful of resources, depending
upon the distribution of future requests. In general, the SL-based strategy can be
effective if the \epss maintained at any particular SL can be useful in generation of 
many $(s,d)$ \epss.
%%%%%%%%
For example, consider a set/cluster of sources $S$
and a cluster of destination $D$. Let us say that, periodically, we need to serve an \eps request 
over some $(s_i, d_i$) pair where $s_i \in S$ and $d_j \in D$. For such an \eps demand model,
just a single super-link $(A,B)$ could be very useful wherein $A$ is close to all the nodes in $S$
and $B$ is close to all the nodes in $D$. Essentially, an \eps over $(A,B)$ can be used to 
efficiently generate an \eps over any $(s_i, d_j)$ pair where $s_i \in S$ and $d_j \in D$. See Fig.~\ref{fig:sl-cluster}.
%%%%%%%%
In general, for a given \epss demand distribution, there is a need to determine a efficient set of SLs 
that will be most effective in serving the given demand distribution. We formulate this optimization
problem in the next subsection.

\subsection{\bf \sls Problem Formulation}
\label{sec:problem}

In this subsection, we formalize the {\em selection of super-links (\sls)} optimization problem, which is to efficiently select super-links (SLs) in a given 
network \epss to effectively handle a given demand distribution of \epss in the network. 
We start with describing the overall setting.

\para{Problem Setting.} 
Consider a quantum network and a set of source-destination pairs $S = \{(s,d)\}$. 
Let us assume the time to be divided in slots of size $\tau$, and that at the start of each time
slot, there is a single \eps request for one of the $(s,d)$ pairs in $S$.\footnote{The derivation of
Eqns.~\eqref{eqn:sl-latency-1} and~\eqref{eqn:sl-latency-2} implicitly assume that $\tau$ is large enough
that the \epss for all the SLs are available when a new $(s,d)$ \eps request arrives, i.e., 
$\tau$ is larger than the maximum latency incurred in generating the desired number of 
\epss over an SL in the network. Note that \epss across SLs are generated independently and in 
parallel, as they use disjoint paths for \eps generation.}
%%%%%%%%%%%%%%
In order to serve these requests with minimum latency, we select a set of 
SLs based on the below formulated optimization problem, 
and continuously generate \epss over these SLs.
%%%%%%%%%%%%%%
When an $(s,d)$ \eps request is received, an \eps over $(s,d)$  is generated using 
an efficient swapping-tree using already-available \epss at one of the SLs (say $l$). Also, 
while the request is being served, we stop generating \epss over any SL other than $l$ whose
associated path intersects with the $(s,d)$'s \eps-generation path (to allow for most efficient
generation of \eps over $(s,d)$). 
%%%%%%%%%%%%%%%%%%%%%
%As the Sl \eps are consumed or decohere, they are regenerated (continuously).
%%%%
Throughout the paper, we implicitly assume that we use at most one 
SL in generating an \eps for an $(s,d)$ pair. 
The main motivation behind this assumption to limit the number of \epss required at any SL (e.g., if we use
two SLs in an $(s,d)$ swapping tree, then the number of \epss required would be $1/\bp^4$). 
%%%%%%%%%%%%%%%%%%
Before formulating the optimization problem addressed here, we define two functions.

\softpara{Objective Function $\psi$.} 
Consider a set $S$ of $(s,d)$ pairs and a set of super-links $L$. 
We define a function $\psi(S,L)$ as the average generation latency of the pairs 
in $S$ using the super-links in $L$. In particular, let $T((s,d), l)$ denote the
expected latency in generating an \eps pair over $(s,d)$ using a super-link $l$; this
is essentially a minimum of the two qualities given in Eqns.~\eqref{eqn:sl-latency-1}-\eqref{eqn:sl-latency-2} 
where $l = (A,B)$. Then, the function $\psi(S,L)$ is essentially given by: 
$$\psi(S,L) = \sum_{(s,d) \in S}   \min_{l \in L} T((s,d), l).$$

\softpara{Cost Function: $Cost(L)$.}
To capture efficient utilization of network resources, we now model the cost of super-links. There are many 
ways to model the cost---but in general, the idea is for the cost to reflect the network resources consumed in generation of \epss over the set of SLs. One simple way to measure the network resources consumed in 
generating an \eps could be just the latency incurred in generating it---and based on this, the cost of
an SL could be defined as the expected generation latency of an \eps over the SL using the associated swapping-tree.
%%%%%%%
In our discussion, however, we capture the network resources consumed more accurately by modelling it in terms of the aggregate number of \eps generation attempts made by the links in the path (which are also the leaves of the associated swapping-tree) associated with the super-link, in generating an \eps at the SL. In other words, the cost of an SL $l$ is the number of BSM failures incurred at the link-level of the swapping tree associated with the SL $l$, in generation of an \eps at the SL.\footnote{In reality, we generate multiple \epss at each SL, but since this number is uniform is across all SLs, the cost can be based on generation of a single \eps.}
%%%%%%%%%%%%%%%%%%%%%%%
Thus, the cost of a set $L$ is just the sum of the costs of the individual super-links. 
%%%%%%%%%%%%%%%%
We note that the techniques developed in this work is independent of the cost model for super-links. 

\para{Selection of Super-Links (\sls) Problem.}
Given a quantum network and a set $S$ of $(s,d)$ pairs, the \sls problem is to select a set $L$ of 
super-links that minimizes $\psi(S, L)$ such the total cost of the super-links $L$ is at most a given budget. 
\softpara{Hardness and Variants.}
The \sls problem can be easily shown to be NP-complete, from a simple reduction from the well-known
set cover problem. One could look at other variants of the problem -- e.g., minimize the maximum latency
of the $(s,d)$ pairs, minimize the cost of the SLs under the constraint that the latency of each $(s,d)$
pair is bounded by a given constant, etc. The algorithms designed in our paper can be modified to address
these variants. 

% \begin{figure}
%     \centering
%     \includegraphics[width=0.4\textwidth]{conext-submission/figures/cluster1.PNG}
%     \vspace*{2in}
%     \caption{\sls Example. Clustered Pairs and Super-Links.}
%     \label{fig:sls-cluster}
% \end{figure}

% \begin{figure}
%     \centering
%     \includegraphics[width=0.4\textwidth]{conext-submission/figures/cluster2.PNG}
%     \vspace*{2in}
%     \caption{\sls Example. Clustered Pairs and Super-Links.}
%     \label{fig:sls-cluster}
% \end{figure}

% \para{\sls Example.}
% \red{This is important. Show (S,D) example. (s,d) example. Show clustered images.}
% \vspace{2in}

\subsection{\bf Related Works.}
\label{sec:related}

There have been a few works in the recent years that have addressed generating long-distance \epss
efficiently. These works have focused on selecting an efficient routing path~\cite{caleffi, sigcomm20, delft-lp} or swapping-tree~\cite{swapping-tqe22} for the swapping process.
The closest work in the quantum communication literature to ours is~\cite{greedy2019distributed}, which
shows that use of pre-distributed entanglements can reduce the latency of generating entanglements, and 
develops routing algorithms in special network topologies (e.g., rings, grids) that leverage these pre-distributed entanglements; we note that~\cite{greedy2019distributed} refers to the our super-links as
virtual links.
%%%%%%%
Our work focuses on the related problem of selection of links where these pre-distributed entanglements should
be generated---given a distribution of future entanglement requests.
%%%%%%%%%%%%%%%%%%%%%%%
In addition to the above works, other works that have focused on developing quantum networking protocols 
include~\cite{sigcomm19} which develops a link-layer protocol for entanglement generations,~\cite{conext20} which develops a network protocol for efficient use of entanglement-pairs for swapping operations in 
entanglement generation over remote network nodes.

Our proposed optimization problem \sls of selection of super-links has not been 
studied before. The closed optimization problem studied before is that of {\em adding 
short-cuts in a graph (\adsg)} by Myerson~\cite{myerson-2009}. The \adsg problem addressed
in~\cite{myerson-2009} proposes to add a certain number of short-cuts to a given graph with 
the objective of minimizing the average path length across all pairs of nodes. They propose
approximation algorithms for the above graph problem based on known approximation algorithm
for the $k$-medians optimization problem. The short-cuts in our \sls problem can be looked
upon as short cuts in the \adsg problem. For the \adsg problem,~\cite{myerson-2009} consider
both versions: allowing multiple short-cuts in a path, or only a single short-cut for each
path.
However,  the fundamental differences of our 
\sls problem with the above \adsg problems are: (i) In \adsg, the short-cuts are assumed to be
of uniform weight (they use a certain {\em number} of short-cuts; generalization of their techniques to weighted short-cuts requires fundamentally new techniques, because the weighted version of $k$-medians problem itself has not been studied before to the best of our knowledge.
(ii) The short-cuts in \adsg are disjoint and thus independent by definition, while in our \sls problem, the biggest challenge arises from the inter-dependence been the super-links (see \S\ref{sec:greedy}. 
%%%%%%%%%
In addition to the above, other related work includes~\cite{sk-99} which addresses a problem related to the dual version of the \sls problem; in particular,~\cite{sk-99} look at
{\em mincost distance-$d$} problem which attempts to bound the distance between all pairs
of nodes to $d$ by adding short-cuts of total minimum cost. Here, the cost of the short-cuts is non-uniform. The {\em mincost distance-$d$} is related to the dual of the \sls problem, wherein
we want to add super-links of total minimum cost such that the \eps generation latency of each
$(s,d)$-pair is bounded by a certain constant; however,~\cite{sk-99}'s notion of distance between pairs of nodes is much simpler compared to the notion of \eps generation latency, and the techniques in~\cite{sk-99} do not generalize to the \sls-dual problem (and certainly, not to the
\sls problem addressed here).

%They gives an O(n log d) approximation algorithm. However, diameter concept is very different than latency -- and their techniques don't generalize. Moreover, they look at a more general version -- allowing multiple SLs per pair. They show the equivalence of mincost diameter-d” === Mincost k-spanner (complete graph), use ILP, LP relaxation and round up. Bilo-2012 improves the results for the special case when SL-costs are uniform. 

% \item
% ~\cite{myerson-2009} look at a similar problem -- but focuses on the objective of minimizing the sum of path lengths using at most k SLs.~\cite{myerson-2009} gives approximation algorithms based on k-median problems (assumes “metric” cost). Somehow the "dual" of our problem--but for multiple SLs/path and assumes metric cost. 

% \item
% ~\cite{khanna-99} shows  that “mincost diameter-d” === “Mincost k-spanner” problem. Results on mincost k-spanner problem:
% Greedy seems to do well. Best Analysis.
% Best algorithm (only slightly better than greedy). These suggest the value of greedy algorithms. 
% \end{itemize}
% Key points:
% (i) We use a simplified version -- of allowing only SL/path, which makes it much more tractable. 
% (ii)  For multiple SLs, our problem -- has a unique latency (and cost?) model, which makes most techniques and prior analysis inapplicable. We instead use a modified version of single-SLs.
% (iii) For additive problem, single SL again makes it easy. Multiple SLs .. is tricky .. (no known approach applies).
% \newpage

\section{\bf Greedy Algorithm} 
\label{sec:greedy}

In this section, we design a greedy algorithm for selection of SLs. At a high-level, a greedy approach for the \sls problem would iteratively select an SL based on some
criteria, until the given budget is exhausted. For optimization problems with a
submodular objective function, an appropriately designed greedy
approach can be shown to deliver a solution whose objective value is within a constant-factor of the optimal objective value, i.e., a constant-factor approximate
solution. Unfortunately, the \sls problem's objective function is not submodular, and
hence, a greedy approximation algorithm is not feasible. Nevertheless, an appropriately
designed greedy algorithm can be expected to deliver good solutions in practice. 
%%%%%%%%%%%%%
Below, we start by motivating designing a greedy approach for the \sls problem, by showing that for a special case of the \sls problem, a simple greedy algorithm delivers a constant-factor approximate solution. With
this motivation and insight, we design an appropriate greedy algorithm for the
general \sls problem; the designed greedy algorithm delivers good solutions in practice as shown in our empirical results (\S\ref{sec:eval}). 

\para{\gdsls Special Case: Given Disjoint SL Candidates.}
Consider a special case of the \sls problem, wherein we are {\em given} a set 
$L$ of {\em disjoint} 
paths in the network and the problem is to select a subset 
$L' \subseteq L$ of SLs that yields the
minimum aggregate latency of the $(s,d)$ pairs under the cost constraint. 
%%%%%%%%%%%%%%%%%%%
We refer to this special case as the \gdsls problem. 
%%%%%%%%%%%%
For this special case, a straightforward greedy approach of iteratively 
selecting the SL that has the maximum  "benefit" per unit cost can be 
shown to deliver
a constant-factor approximate solution. 
%%%%%%%%
The benefit of an SL $l$ is defined as the reduction in the aggregate 
latency of $(s,d)$ pairs
due to adding $l$ as an SL; we formalize this notion of benefit below.

\softpara{Benefit of an SL.}
Consider a stage of the greedy algorithm where a set of
SLs $L'$ have already been selected. At this stage, the {\em benefit} of 
an SL $l \notin L'$ is
defined as $$\psi(L') - \psi(L' \cup\ \{l\}),$$ where the $\psi$ function, 
as defined earlier, is the aggregate latency of the given $(s,d)$ pairs 
using the set of SLs.

We now give the pseudo-code of the proposed greedy approach for the \gdsls problem. 

\begin{algorithm}
    \caption{Greedy for \gdsls Problem.} 
    \begin{algorithmic}[1]
    \State \textbf{Input:} Quantum Network, $\{(s,d)\}$ pairs, $L$ of candidate SLs, Cost Budget $C$. 
    \State \textbf{Output:} Set $L'$ of Super-Links.
    \State $L' = \phi$
    \While{$({\rm cost}(L') < C)$ and ($L-L' \neq \phi$)} 
        \State Let $l$ be the SL in $L - L'$ with the highest value of $$(\psi(L') - \psi(L' \cup {l}))/Cost(l)$$ 
        \State $L' = L' \cup \{l\}$
    \EndWhile   
    \State \textbf{Return} $L'$
    \end{algorithmic}
\end{algorithm}

The aggregate latency function can be easily shown to be monotonic and submodular, in the context
of the special-case of \gdsls. Thus, the Greedy algorithm can be shown to deliver a solution with a near-optimal benefit.
%\vspace*{-0.1in}
\begin{theorem}
For the special case \gdsls problem, the Greedy algorithm delivers a solution that has a benefit of at least $63\%$ of the benefit of the optimal solution.
\end{theorem}

\begin{figure}
\vspace*{-0.2in}
    \centering
    \includegraphics[width=0.4\textwidth]{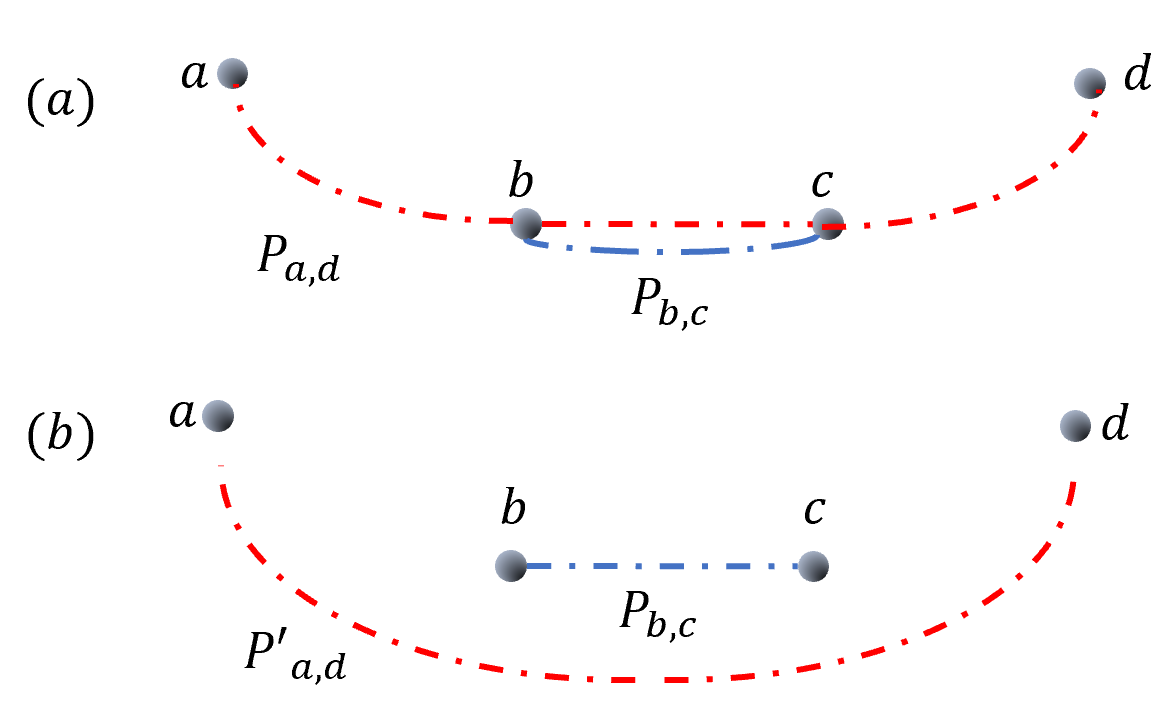}
    % \vspace*{2in}
    \caption{Illustrating the simple greedy scheme, and the \geng Algorithm with {\em Update} and {\em Append} options. (a) Two SLs $(a,d)$ and $(b,c)$ shown with associated (shortest) paths $P_{ad}$ and $P_{bc}$ respectively; these paths are not disjoint. (b) New non-shortest path $P'_{ad}$ for $(a,d)$ which doesn't intersect with $(b,c)$'s associated path $P_{bc}$. 
    %%%%%%%%%%%%%%%%%%%%%%%%%%%%%%
    If SL $(b,c)$ (with associated path
    $P_{bc}$) is first/already chosen as an SL in the solution, then a {\em simple greedy algorithm} will not consider $(a,d)$ as a super-link since its associated path $P_{ad}$ intersection with $P_{bc}$.
    Instead, the \geng Algorithm considers including super-link $(a,d)$ in the solution via two options: (i) {\em Update:} Remove $(b,c)$, and add $(a,d)$ with associated (shortest) path $P_{ad}$, or (ii) {\em Append:} Add $(a,d)$ with associated path $P'_{ad}$.}
    \label{fig:shortest_vs_non}
    %\vspace*{-0.2in}
\end{figure}

\para{Generalized Greedy (\geng) for the \sls Problem.} Note in the general \sls problem, there are no given candidates; in fact, each path in the network can be an SL -- so, the number of candidate SLs is exponential in the number of network nodes. 
%%%%%%%%%%%%%%
However, it is reasonable to consider only the best (i.e., ones with lowest-latency swapping trees)
path for every pair of nodes in the network; this yields at most $O(n^2)$ candidate SLs, where $n$ is the number of nodes in the network. Let us use \po to denote this set of candidate SLs.
%%%%%%%%%%%%
The set of candidate SLs in \po might however be non-disjoint, i.e., have a common node or link. Intersecting candidate SLs creates multiple issues: (i) The objective function is not submodular for
SLs with intersections; (ii) Allowing non-disjoint SLs requires appropriate allocation of node and link resources across SLs to generate \eps for the SLs, and more importantly, would results in higher generation latencies for the \eps over the SLs.
%%%%%%%%%%%%%%%%%%%%%
For the above reasons and also for simplicity, we enforce the condition that the selected SLs be disjoint.
%%%%%%%%%%%%%%%%%%%
With that condition in mind, 
a straightforward greedy approach for the general \sls problem could be to iteratively select the SL (from the candidate set \po) with the highest benefit per unit cost among those that do not intersect with the already selected SLs. 
%%%%%%%%%%%%%%%%%%
This simple approach can easily perform badly (see Fig.~\ref{fig:shortest_vs_non}), 
but can be further improved in one of the following options: 
(i) Allow for non-shortest paths not in \po, if needed; (ii) Allow removal of previously selected SLs if they intersect with a promising candidate later. 
%%%%%%%
In our greedy algorithm for the general \sls problem, which we refer to as \geng, 
we incorporate both the above options. We present \geng's pseudo-code as 
Algorithm~\ref{alg:general_greedy} below. In \geng, we essentially evaluate each pair of nodes $(u,v)$ 
as a potential SL for addition to the set of SLs $S$ being maintained. 
%%%%%%%%
We consider two options as follows (see Fig.~\ref{fig:shortest_vs_non}): 
(i) {\em Update:} If we allow deletions from $S$, then we pick the shortest path
$P_U$ from $u$ to $v$ in the original graph, delete all the SLs from $S$ whose associated paths intersect with $P_U$. (ii) {\em Append:} If we do not allow deletions from $S$, then we instead pick the shortest path $P_A$ from $u$ to $v$ in the ``residtual'' graph to avoid intersections with associated paths of SLs in $S$. 
%%%%%%%
For each potential option (two for each pair of nodes), we compute the ratio of increase in benefit to increase in cost, and pick the option with the best ratio. We iterative the above, until the given cost budget is exhausted.
%%%%%

% \blue{2-3 sentence description of the psuedo-code/algorithm.}
% (\red{example, see Fig.~\ref{fig:shortest_vs_non} where two \sls $sl_1$ and $sl_2$ are the best to serve $(s_1, d_1)$ and $(s_2, d_2)$ respectively. As it is shown in Fig.~\ref{fig:shortest_vs_non}, selection of one $SL$ invalidate the selection of the other which leads to a higher overall latency. On the other hand, if we select a non-shortest path for $sl_2$, which has no intersection with $sl_1$, we can get the performance of selecting both \sls independently.})

% \begin{figure}
%     \centering
%     \includegraphics[width=0.4\textwidth]{conext-submission/figures/non_shortest_vs_shortest.PNG}
%     % \vspace*{2in}
%     \caption{In this example, if $sl_1$ is chosen first, then .}
%     \label{fig:counter-example}
% \end{figure}

\begin{algorithm}
    \caption{Generalized Greedy (\geng) Algorithm for \sls} 
    \label{alg:general_greedy}
    \begin{algorithmic}[1]
    \State \textbf{Input:} Quantum Network $G$, $\{(s,d)\}$ pairs, Cost Budget $C$
    \State \textbf{Output:} Set $L$ of Super-Links.\footnotemark  \footnotemark
 % \State AllPairsShortestPaths($G$) be a sub-module where shortest paths between every pair in a network, represented by $G$, arecalculated.
 %   \State SLAssignement($S$, $L$) be asub-module where tries to assign a $l\in L$ to a $s\in S$ s.t. the final latency of $s$ is minimum.
 %   \State ResidualGraph($P$) be a sub-module where the residual graph by subtracting all resources used by the final paths (including $L$) is calculated.
 %   \State DisjointSLs($L$, $l$) be a sub-module where the maximal \textit{DISJOINT} of $L$ in regard to $l$ is returned (plus $l$) (deletion of $L$ may happen).
 
    % \State $M$ = All pair of nodes in $G(V, E)$ s.t. every pair of nodes $m\in M$ is able to generate $1/(\bp)^2$ of \eps in time period of \demandT.\footnote{some text}
    \State $L = \phi$
    \While{${\rm cost}(L) < C)$}  
       
        % \State $M' =$ AllPairsShortestPaths($G'$)
        \For{$ (u,v) \in ((V \times V) - L)$}
        
            /* {\em Update}: If added with deletions from $L$ */
            \State $P_U=$  shortest path between $(u,v)$ in $G$
            \State $X =$ set of SLs in $L$ whose associated paths are disjoint with  $P_U$.
            \State $L_U' = X \cup \{(u,v)\}$
            \State $b_U=\big(\psi(L) - \psi(L_U')\big)/(Cost(L_U') - Cost(L)$ \footnotemark
                 
            /* {\em Append}: If added without any deletions from $L$ */
            \State $P_A=$ shortest path between $(u,v)$ in the ``residual'' graph $G'$ where $G$' is $G$ without the nodes and edges used in the paths associated with SLs in $L$).
            \State $L_A' = L \cup \{(u,v)\}$
            \State $b_A=\big(\psi(L) - \psi(L_A')\big)/Cost(\{(u, v)\}))$

            \State \If{$b_A > b_U$}
                \State $b_{u,v} = b_A$
                \State $L'_{u,v} = L'_A$
            \Else 
                \State $b_{u,v} = b_U$
                \State $L'_{u,v} = L'_U$
            \EndIf
        \EndFor
        \State $L' \gets L'_{u,v}$ with maximum $b_{u, v}$ value.
        \If{$L=L'$}
            \State break
        \EndIf
        \State $L = L'$ 
    \EndWhile 
    \State \textbf{Return} $L$
    \end{algorithmic}
\end{algorithm}
 %\vspace*{-0.1in}
 
% \para{Remarks.}
% \blue{No performance guarantee -- mainly because the SLs in \P may intersect. 
% There is chance to develop an upper bound: Make the candidates in \P disjoint, and do a greedy. Optimal cannot be better than 1/0.63 of the greedy solution. Needs confirmation.}

\section{\bf Clustering-Based Algorithm}
\label{sec:clus}

In this section, we 
design an algorithm based on clustering the $(s,d)$ pairs and picking an SL for each
of the clusters.
%of $(s,d)$ pairs. We start with the intuition and basic idea. 

\para{Intuition and Basic Idea.}
Intuitively, the idea is to pick SLs each of which can 
effectively serve as many $(s,d)$ pairs as possible, 
to allow for 
efficient use of resources. 
Therefore, given a set of $(s,d)$ pairs, we want to partition the set into a small number of clusters  in a way that each cluster of $(s,d)$ pairs can be effectively served by an appropriately picked 
single SL. 
%%%%%%%%%%%%%%%%%%%%
%%%%%%%%%%%%%%%%%%%%%%%%%%%%%%%%%%%%
We can use a clustering algorithm based on the standard $k$-means clustering algorithm.
%%%%%%
For a given $k$, the $k$-means algorithm starts with a randomly chosen $k$ cluster-centroids (SLs in our case) and iteratively (i) assigns each $(s,d)$ pair to the "closest" centroid SL; the set of $(s,d)$ pairs assigned to the same centroid SL forms a cluster; (ii) computes new centroid SLs for the clusters formed; (iii) repeat until some condition is
satisfied. 
%%%%%%%%%%%%%%%%%
Since in our \sls problem, the constraint is in terms of the total cost budget rather than the number of clusters $k$, we iterate over potential $k$ values and pick the best clustering solution that satisfies the budget constraints. 
%%%
Even for a given $k$, to use the above $k$-means approach, we still need to define/determine: (a) The distance between an
SL and an $(s,d)$ pair to determine the closest SL for centroid assignment, and (b) How to compute a centroid for a given set of $(s,d)$ pairs. We define these aspects as follows. 
\begin{itemize}
    \item {\em Distance between SL and an $(s,d)$ pair.} The distance between  an $(s,d)$ pair and a SL can be defined as the latency of the $(s,d)$ pair using the SL. Thus, all the $(s,d)$ pairs in the same cluster would have the common property that the centroid SL of this cluster reduces their latency the most compared to other centroid SLs. 

    \item {\em Centroid SL of a set of $(s,d)$ Pairs.} We use an exhaustive way to find the updated centroids for each cluster. In particular, for each cluster $C_{i}$, we assume any of the paths $p$ in the network can be the new centroid, and  traverse through all the paths in the network and pick the one whose aggregate distance (as defined above) from the $(s,d)$ pairs in the cluster is minimum.
\end{itemize}
We now present the overall pseudo code below. 
\addtocounter{footnote}{-3}
\stepcounter{footnote}\footnotetext{As mentioned before, although a super-link is defined by a pair of nodes, it is also associated with a path used for the swapping-tree.}
\stepcounter{footnote}\footnotetext{A super-link is  considered only if its associated path is able to generate $1/(\bp)^2$ \epss with an expected latency of less than \demandT.} 
\stepcounter{footnote}\footnotetext{The special case is when the denominator is negative while the nominator is positive; such an SL is considered better than those with positive $b$ values.}
\addtocounter{footnote}{-1}

\begin{algorithm}
    \caption{Clustering Algorithm for \sls Problem} 
    \begin{algorithmic}[1]
    \State \textbf{Input:} Quantum Network $G$, S = $\{(s,d)\}$, Cost Budget $C$.
    \State \textbf{Output:}
    \For{$k = 1, 2,\ldots, |S|$ }
        \State $L_k = $ ClusteringViaKMeans$(G, S, k)$
    \EndFor
    \State \textbf{Return} $L_k$ where $k=\argmin_{k=1,2,..}\psi(L_k)$
    \end{algorithmic}
\end{algorithm}
    % \If{All solution are ignored}
    %     \State Pick the solution with min unsatisfied (s,d) pairs
    %     \State $\forall$ unsatisfied pairs, add min cost $SLs$ to satisfy them
    % \EndIf
    % \State \textbf{Return} non-ignored min $TotalCost$ / extended solution

\begin{algorithm}
    \label{alg:clustering}
	\caption{Selecting $k$ Super-Links via $k$-means Clustering} 
	\begin{algorithmic}[1]
	\State \textbf{Function Name:} ClusteringViaKMeans
	\State \textbf{Input:} k, Quantum Network $G$, $S = \{(s,d)\}$ pairs, Cost Budget $C$.
	\State\textbf{Output:} Set $L$ of Super-Links (also, centroids of the $k$ clusters).
%	\State $SLs$ = AllPairShortestPaths($G$) s.t. every pair of nodes $m\in M$ is able to generate $1/(\bp)^2$ of \eps in time period of \demandT.
%	\State Randomly partition $S$ into $k$ clusters/subsets $S_1, S_2, \ldots, S_k$. 
	\State Randomly pick $k$ SLs/centroids $l_1, l_2, \ldots, l_k$.
	\While{()}
	    \State /* Assignment Stage */
	    \For{each pair $(s, d) \in S$}
	        \State Find the SL $l_j$ with minimum latency $\psi((s, d), l_j)$
	        \State Assign $(s, d)$ to the SL/centroid $l_j$.
	    \EndFor
	    \State /* Update SLs/Centroids */
	    \For{$1 \leq i  \leq k$}
	        \State Let $S_i$ be the set of $(s,d)$ pairs assigned to SL $l_i$. 
	        \State Let $l$ be the super-link\footnote{We only consider SLs that have a swapping-tree of latency less than the \blue{demand period} \demandT.} (with an associated path and swapping-tree) with minimum $\psi(S_i, l)$, .
	        \State $l_i = l$
	    \EndFor
	    \State /* Determine whether to break. */
	    \State Keep track of the set of SLs with best $\psi(S, \{l_1, l_2, \ldots, l_k))$, and 
	    {\bf break} if the best set of SLs doesn't change in 5 iterations of the while loop.
	\EndWhile
	\State if $Cost(L) \geq C$ then reduce $Cost(L)$ by ''shortening'' (i.e., using a sub-path) some $l\in L$
	\State \textbf{Return} $\{l_1, l_2, \ldots, l_k)$
	\end{algorithmic} 
\end{algorithm}

%	    $TotalBenefit = \sum_{CL_{i}}\sum_{(s,d) \in CL_{i}}\frac{Latency(s,d) - LatencyWithSL(s, d, L_{i})}{cost(L_i)}$
%	    \State Finish when one's $TotalBenefit$ wins 5 times

            % /* If added without any deletions from L */
            % \State $p_{{u, v}_A}=$ shortest path between $(u,v)$ in the ``residual'' graph (i.e., $G$ without the edges used in the paths associated with SLs in $L$).
            % \State $L_{{u,v}_A}' = L \cup \{(u,v)\}$
            % \State $b_{{u, v}_A}=\big(\psi(L) - \psi(L_{{u,v}_A}')\big)/{\rm cost}(u, v)$

            % /* If added with possible deletions from L */
            % \State $p_{{u, v}_U}=$  shortest path between $(u,v)$ in $G$
            % \State $X =$ set of SLs in $L$ whose associated paths are \red{edge-disjoint} with  $p_{{u, v}_U}$.
            % \State $L_{{u,v}_U}' = X \cup \{(u,v)\}$
            % \State $b_{{u, v}_U}=\big(\psi(L) - \psi(L_{{u,v}_U}')\big)/({\rm cost}(L_{{u,v}_U}') - cost(L)$ \footnotemark
            
            % \State \If{$b_{{u, v}_A} > b_{{u, v}_U}$}
            %     \State $b_{u,v} = b_{{u, v}_A}$
            %     \State $L'_{u,v} = L'_{{u, v}_A}$
            % \Else 
            %     \State $b_{u,v} = b_{{u, v}_U}$
            %     \State $L'_{u,v} = L'_{{u, v}_U}$
\section{\bf Network Protocol and Implementation}
\label{sec:protocol}

In this section, we present the network protocol with other implementation details, used 
to generate \epss over selected SLs and then, to use these pre-distributed \epss over SLs to serve incoming communication requests.  
%%%%%%
Overall, the network protocol has the following high-level components/steps: 
\begin{itemize}
    \item {\bf Selection of SLs.} This step is largely done offline at a centralized (classical) node, with the inputs being the \eps requests (i.e., the weighted set of $(s,d)$ pairs) and the quantum network details. However, as the request distribution changes, the set of SLs are updated. 
    
    \item {\bf Continuously Generate \epss across Selected SLs.} We continuously generate \epss over the selected SLs, using all the link resources available (note that our selected SLs are disjoint), so as to keep them as ``fresh'' as possible. As we use at most one SL per swapping-tree of a future $(s,d)$ \eps request, we store  $1/\bp^2$ \epss at each SL since that is the expected number of \epss needed at an SL. In addition, we also keep track of the generation time of the qubits in each \eps pair, so that we can appropriately determine if they can be used. Generation of \epss across the SLs is done using the swapping-tree protocol described  below.
    
    \item {\bf Serve Communication Requests.} Note that for any $(s,d)$ \eps request, we already know the SL that we will use in its swapping-tree---since, such assignment of SLs to $(s,d)$ pairs is implicitly already done during selection of SLs.\footnote{In a more sophisticated scheme, we could determine the SL to be used by an $(s,d)$ pair at real-time based on the current ages of the \epss available at the SLs. We will explore this generalization in our future work.} For a given $(s,d)$ pair request, let $(x,y)$ be the SL used. Now, consider the sequence of nodes (i.e., the entanglement path) $\langle s, \ldots, x, y, \ldots, d\rangle,$ over which the swapping-tree is constructed. 
    %%%%%%%%%%%%%%%%%%%
    We discuss generation of \eps across $(s,d)$ using such a swapping-tree involving the SL $(x,y)$ below. Note that we need to only generate one \eps (see~\S\ref{sec:gen} for the general case).
\end{itemize}

%%% Assume classical connections. 
%%% All nodes know their role within the SL swapping tree. We hide details. 

\para{\eps Generation over a Given Swapping Tree.}
Consider a a swapping tree $T$ for a path $P$ connecting a pair of nodes $(x,y)$
in the network. We build our protocol on top of the link-layer of~\cite{sigcomm19},
which is delegated with the task of generating \epss on a link at a desired rate.
%%%%%%%%%%%%%%%%%%%%%%%%%%%%%%%%%%%%%%%%
In our case, we select disjoint SLs---thus, each link of on SL 
is involved in a single SL swapping-tree;\footnote{If a link $(a,b)$ is part of multiple swapping-trees
of different SLs, it needs to handle multiple link-layer requests at the same time. Such link-layer
requests can be implemented by creating ``virtual'' independent atom-photon generators at $a$ and $b$, with one pair of synchronized generators for each link-layer request. See~\cite{swapping-tqe22} for 
more details.} Thus, each link in an SL can continuously generate \epss at the 
maximum rate possible, unless instructed otherwise (see below).
%%%%%%%%%
As mentioned before, for a given SL (and a corresponding path), we compute the optimal swapping-tree using techniques in~\cite{swapping-tqe22}. The network protocol that
generates \epss over a {\em
given} swapping tree can be described as follows.
%%%%%%%%%%%%%%
For simplicity, let us first consider generation of a {\em single} \eps at a time; we 
consider multiple \epss shortly.
The key aspect of the protocol is that the entanglement-swapping (ES)  operation is done only when both the 
appropriate \eps pairs have arrived. The key step of the protocol is the bell-state measurement
(BSM) or the ES operation done at appropriate nodes on the path. Consider a pair of \epss 
over $(A,B)$ and $(B,C)$, with the node $B$ now planning to do an ES operation to generate
an \eps over $(A,C)$; this situation corresponds to siblings $(A,B)$ and $(B,C)$ and their
parent $(A,C)$ in the swapping tree. This BSM operation at $B$ essentially entails the following:
%%%%%%%%%%%%%%%%%%
\begin{itemize}
    \item If the swapping (BSM) operation at node $B$ succeeds, it transmits classical bits to $C$ which manipulates its qubit, and send the final ack to the other end-node $A$ to finalize ``generation'' of the \eps over $(A,C)$. 
    
    \item If the swapping (BSM) operation fails, a classical \ack is send to the end nodes $A$ and $C$ to discard the corresponding qubits. 
    %link-leaves of the subtree rooted at $(A,C)$ of the swapping tree. This \ack is used to notify the links to start generating new link-\epss; in our protocol, a link $l$ does not accept any more \epss, while its ancestor is waiting for its sibling's \eps. \red{Why?}
\end{itemize}
\softpara{Generating Multiple \epss.} The above discussion is to generate a single \eps at a time. 
%%%%%%%%%%%
To generate multiple \epss, we can do the above simultaneously--not necessarily independently. Essentially, each node in the swapping tree (including the link leaves) stores the \epss that are valid, i.e., that are part of an \eps and haven't 
decohered. The node $B$ (as define above) will initiate an ES operation as soon as $(A,C)$ as well as $(C,B)$ has at least
one \eps each. Note that we can even use different swapping-trees for multiple \epss, even over the same path.

\begin{figure}[t]
    \vspace{-0.1in}
    \centering
    \includegraphics[width=0.5\textwidth]{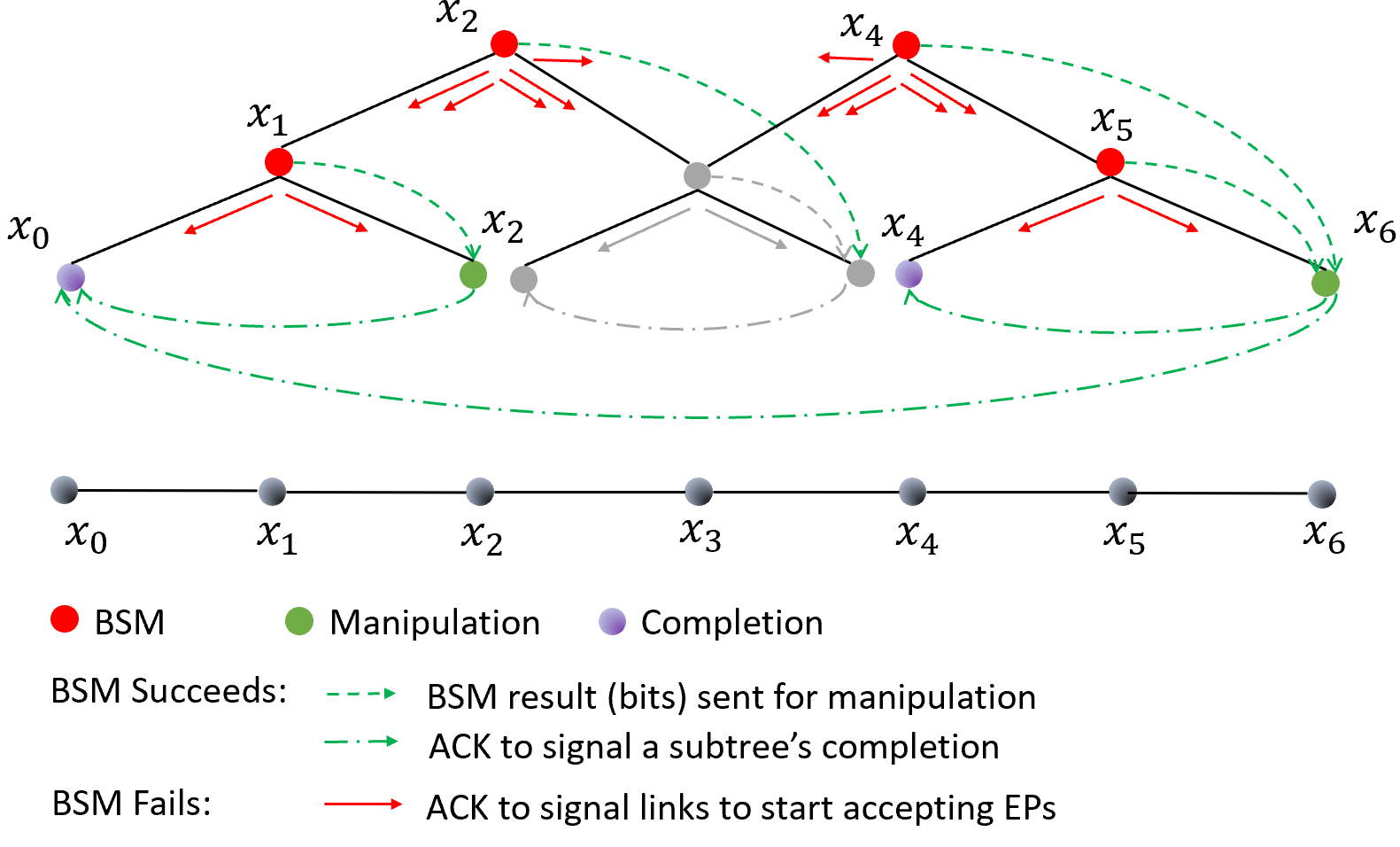}
    \vspace{-0.2in}
    \caption{Serving a communication request over $(x_0, x_6)$, and the ACKs generated following a BSM success or failure.}
    \label{fig:protocol}
    \vspace{-0.2in}
\end{figure}

\para{Serving Communication Requests.}
Consider the sequence of nodes (i.e., the entanglement path) $\langle s, \ldots, x, y, \ldots, d\rangle,$ over which the swapping-tree is constructed; here $(x,y)$ is the SL being used to generate an \eps over $(s,d)$. 
%%%%%%%%%%%
Now, to generate an \eps over $(s,d)$, we essentially use the above protocol as detailed above for a swapping-tree, except for the following: (i) Treat $(x,y)$ as a link with \epss available, (ii) The link  $(x,y)$ continues generating \epss using its swapping tree, independent of the swapping tree used for $\langle s, \ldots, x, y, \ldots, d\rangle.$ (iii) For best performance, during \eps generation of $(s,d)$, we do not use any link on the path $(s,x)$ or $(s,y)$ for generation of any other \epss (i.e., \epss from other selected SLs); other than this constraint, other SLs continue to generate the \epss over them.
%%%%%%%%%
See the example in Fig.~\ref{fig:protocol}, where with the help of a $(x_2, x_4)$ \eps generated from a $(x_2, x_4)$ super-link, a smaller swapping-tree can be used to generate the $(x_0, x_6)$ \eps. 

\section{\bf Generalizations and Discussion}
\label{sec:gen}

In this section, we relax some of the simplifying assumptions made earlier, discuss other variants of the \sls optimization problem, and other issues.

\para{General Distribution of \epss Requests.} Throughout the paper, we have considered a simple model of \eps requests wherein there is a single $(s,d)$ \eps request in each time slot. Our techniques can be easily generalized to handle more general request models: (i) A weighted model, wherein a weight (a fraction) associated with an $(s,d)$ pair signifies the ratio of times a request belongs to this $(s,d)$ pair. For this weighted model, the benefit function can be easily extended by using a weighted benefit from each $(s,d)$ pair, and thus, the greedy
algorithm and its variants can be easily extended. Similarly, the clustering approach also generalized by appropriately extending the centroid of a set of pairs to be based on weighted distance to the $(s,d)$ pairs
in the cluster.
(ii) We can also consider the generalization to multiple $(s,d)$ requests in a timeslot. There are multiples
ways to handle this generalization. We could storing more \epss with each SL, and/or have multiple SLs assigned to each $(s,d)$ pair so that at the time of request one SL is serving exactly one $(s,d)$ pair.

\para{Low SL Replenishment or Decoherence Times.}
We had earlier assumed that the time to replenish the SLs, i.e., the latency incurred to generate the desired \epss over all the selected SLs (this happens in parallel across the SLs, as they are disjoint), is less than the time slot $\tau$. If that is not the case, then, as expected, the benefit of SLs is reduced but can be still significant. In essence, the impact of high replenishment time is that the \epss at SLs may not be readily available when an $(s,d)$ request comes and hence the latency reduction in \eps generation over $(s,d)$ s reduced. Note that \epss over SLs continue to be generated as $(s,d)$ \eps is being generated.

Similarly, lower memory decoherence time reduces the impact of our overall scheme of pre-distributing \epss, but can be minimized by somewhat synchronizing the generation of SLs \epss with that of the $(s,d)$ requests so that -- the SL \epss are as ``fresh'' as possible when needed by the $(s,d)$ request's swapping tree. Note that we do not select SLs whose expected latency of generating \epss is more than the decoherence time.

\para{Other Optimization Objectives.} In this paper, we are focused on the optimization objective of minimizing the aggregate latency of given set of $(s,d)$ pairs under the budget constraint of total cost of super-links selected. However, some related optimization objectives can be equally important. E.g., one can consider the optimization objective of maximizing the number of 
$(s,d)$ pairs whose latency can be lowered to a certain bound $d$ using SLs, under the given budget constraints. This optimization problem can be looked upon as a coverage problem, where an SL covers an $(s,d)$ pair if it helps reduce the latency below the given bound, and appropriate techniques similar to ours can be designed. 
%%%%%%%%%%%%%%
Another variant is to minimize the maximum latency of an $(s,d)$ pair, under given budget constraints; this variant can be solved by an iterative application of the previous variant, by trying decreasing $d$ values. 
%%%%%%%%%%%%%%%%
Finally, one can look at the dual variants of the \sls problem, wherein the objective could be to minimize the lost of the SLs selected, while ensuring some bounds/constraints on $(s,d)$ latencies.

%\item Multiple SLs per path, and mulitple paths.  For multiple paths version, we must have designated paths. Generalization of heuristics is easy.
\section{\bf Evaluations}
\label{sec:eval}

\begin{figure*}[t]
    \vspace{-0.4in}
    \centering
    \includegraphics[width=\textwidth]{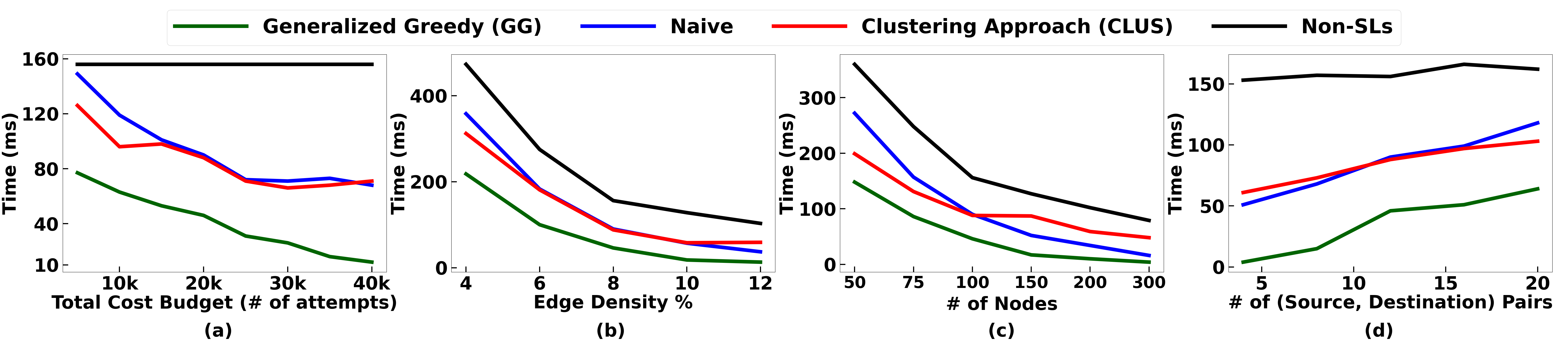}
    \vspace{-0.2in}
    \caption{Average generation latency over $(s,d)$ \eps requests by various algorithms for varying parameters.}
    \label{fig:single_latency}
     \vspace{-0.1in}
\end{figure*}

\begin{figure*}[t]
    \vspace{-0.1in}
    \centering
    \includegraphics[width=\textwidth]{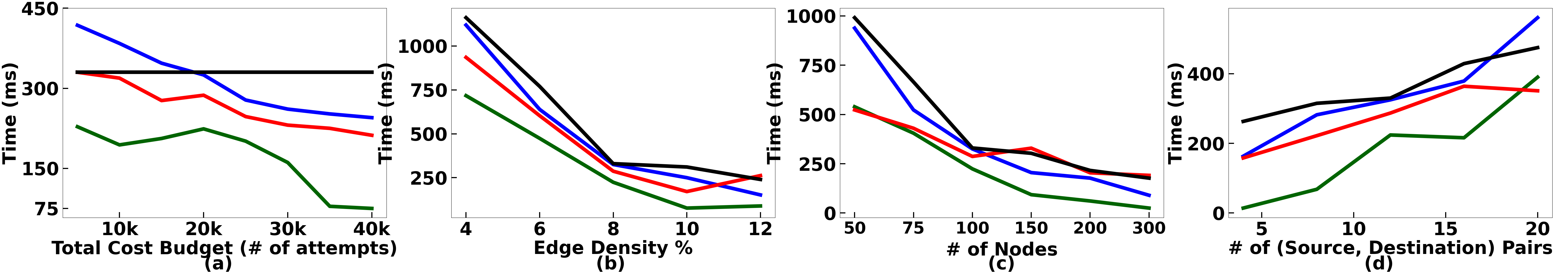}
    \vspace{-0.2in}
    \caption{Maximum generation latency over $(s,d)$ \eps requests by various algorithms for varying parameters.}
    \label{fig:single_max}
    \vspace{-0.1in}
\end{figure*}

\begin{figure*}[t]
    \vspace{-0.1in}
    \centering
    \includegraphics[width=\textwidth]{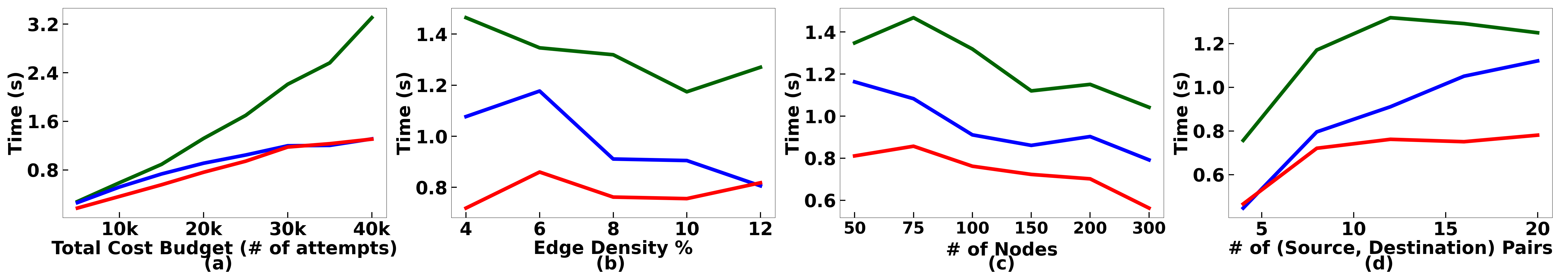}
    \vspace{-0.2in}
    \caption{Aggregate generation latency of super-link \epss incurred by various algorithms for varying parameters.}
    \label{fig:single_sl_latency}
    \vspace{-0.2in}
\end{figure*}

In this section, we evaluate the performance of our protocols/algorithms
in terms of the generation latency of the requesting \epss, by 
implementing and evaluating them on top of the discrete event simulator 
for QNs called NetSquid (Python/C++)~\cite{netsquid2020}.
%%%
The NetSquid simulator accurately models various QN components/aspects, and
%such as 
%physical hardware, fidelity, decoherence, photon transmission, gate/BSM operations,
%and propagation delays.
%%%%%%%%%%%%%%%%%%%%%%%%%%%%
in particular, we are able to (i) 
define various QN components, viz., nodes, quantum memory, 
quantum channels, etc., and (ii) simulate swapping-trees based entanglement 
generation by implementing gate operations needed in entanglement swapping or
teleportation. Our algorithms are implemented in Python. 

\para{Simulation Setting.}
We generate random quantum networks in a similar way as in the recent 
works~\cite{sigcomm20,swapping-tqe22}.
%%%%%%%%%%
By default, we use a network spread over an area of $100 km \times 100 km$.
We use the Waxman model~\cite{waxman}, used to create Internet topologies,
to randomly distribute the nodes and create links; we use the maximum link
distance to be 10km. We vary the number of nodes from 50 to 300, with 100
as the default value. We choose the two parameters in the Waxman model to
maintain the number of links to 8\% of the complete graph (to ensure an 
average degree of 3 to 15 nodes).

\softpara{\eps Generation Parameter Values.}
We use \eps generation parameter values as the 
ones used in~\cite{caleffi,swapping-tqe22}, and vary some
of them. In particular, we use atomic-BSM probability of success (\bp) 
to be 0.4 and latency (\bt) to be 10 $\mu$ secs. The optimal-BSM probability of success (\php) is half of \bp. 
We use atom-photon generation times (\gt) and probability of success
(\gp) as 50 $\mu$sec and 0.33 respectively. Finally, we use photon 
transmission success probability 
as $e^{-d/(2L)}$~\cite{caleffi} where $L$ is the channel attenuation length
(chosen as 20km for an optical fiber) and $d$ is the distance between the nodes.
%%%%%%%%%%%%%%%%%%%%%%%%%%%%%%%%%%%%%
We vary the cost budget (see \S\ref{sec:sls}) used by our algorithms---from $5000$ to $40,000$ units with the default value of $20,000$.

% Each node's memory size is randomly chosen between 15 to 20 units.
% We also provide more memories, $1/\bp^2$ per an SL, for those nodes that are located at either ends of an SL to hold the generated EPs.

\softpara{\epss Request-Traffic Model.} 
We pick the $(s,d)$ pairs such that the distance with $s$ and $d$ is within a range 30 to 120 kms. We vary the number of $(s,d)$ pairs from 4 to 20 with the default value of 12.
%%%%%%%%%%%%%%%%%%%%%%
As described in \S\ref{sec:problem}, we divide the simulation duration into slots of 4 seconds each, and at the start of each slot, there is a request to establish an \eps over one (randomly chosen) of the $(s,d)$ pairs. 
%We also set a constraint on the SLs such that the selected SLs can produce at least $1/\bp^2$ amount of \epss in a Timeslot.}

%%%%%%%%%%%%%%%%%%%%%%%%%%%
% Fidelity is modeled in NetSquid using two parameter values, viz., depolarization
% (for decoherence) and dephasing (for operations-driven) rates. We conservatively 
% choose a depolarization rate of 0.01 such that the fidelity after 1 min
% is 90\%, based on demonstrated coherence times of several 
% minutes~\cite{dec-13,dec-14} to hours~\cite{dec-15,dec-2021}.
% Similarly, we choose a dephasing rate of 1000 which corresponds to a link \eps 
% fidelity of 99.5\%~\cite{delft-lp}.

\begin{figure}
%  \vspace{-0.2in}
    \centering
    \includegraphics[width=0.45\textwidth]{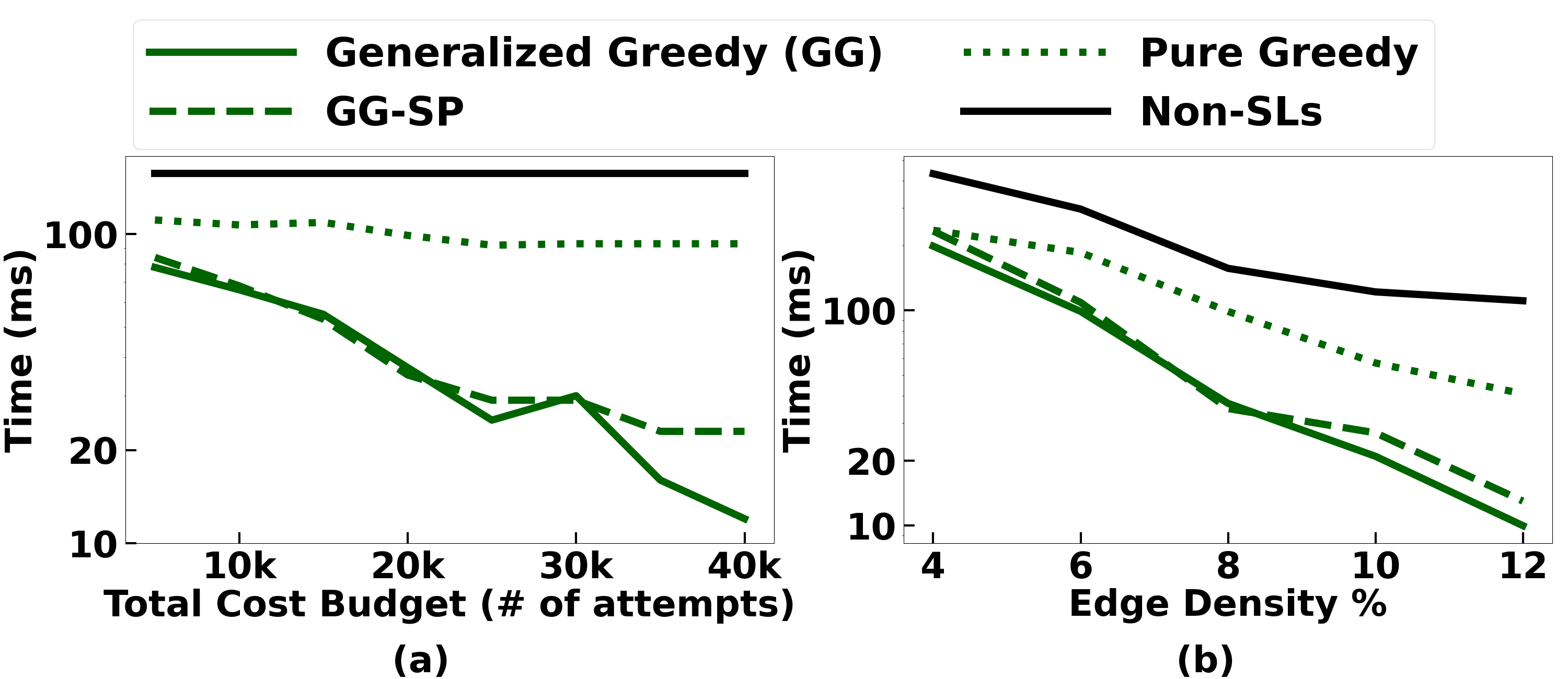}
    \vspace{-0.1in}
    \caption{Average generation latency of $(s,d)$ \eps requests, by certain greedy variants.}
    \vspace{-0.3in}
    \label{fig:signle_greedy}
\end{figure}

\para{Algorithms and Performance Metrics.}
We use the \eps generation algorithm from the latest work~\cite{swapping-tqe22} 
as a benchmark for \eps generation without super-links; the algorithm from~\cite{swapping-tqe22} is based on an dynamic programming algorithm that 
determines an {\em optimal} swapping-tree over a given pair of nodes. We refer to this non-superlink algorithm as \nosl in the plots.
%%%%%%%%%%%%%%%%%%%%%%%%%%%%%%%%%%
For selection of super-links and \eps generation over $(s,d)$ pair with super-links,
we implement and compare the following algorithms: \naive, Generalized Greedy (\geng; \S\ref{sec:greedy}), and Clustering Approach (\clus; \S\ref{sec:clus}). The \naive algorithm is essentially the same as our \geng algorithm, except that the only candidate super-links considered are the sub-paths of the shortest path for a given $(s,d)$ pair.
%%%%%%%%%%%%%%
In addition, we also compare certain variants of the \geng algorithm in one of the experiments. 
%%%%%%%%%%%%
We compare the above algorithms mostly in terms of the average generation latency in seconds of an \eps over an $(s,d)$ pair. In one of the plots, we also evaluate the total latency of the \epss generated over the super-links.
%Finally, we show the cumulative link layer generation rate as another way of showing the cost of SLs picked by different algorithms.
% of the requested (s,d) pairs which are satisfied.

\sloppypar
\para{Evaluation Results.}
We start with comparing the performance of the various \sls algorithms in terms of the average \eps generation latency time to establish an \eps over an $(s,d)$ pair.
%%%%%%%%
See Fig.~\ref{fig:single_latency}, where we plot the average generation latency
for various algorithms for
varying cost budget (over the cost of the super-links selected), network edge density, number of network nodes, and number $(s, d)$ pairs. 
%%%%%%%%%%%%%%%%%%
We observe that all algorithms result in lowered average \eps generation latency, with our main algorithm \geng significantly outperforming all the other algorithms. With enough cost budget (40k units) or with sufficient large or dense network
(e.g. edge density of 12\% or network nodes of 300), \geng can achieve average generation latency as low as 4 milliseconds, while the \nosl algorithm
incurs around 100-200 milliseconds. While the \naive and \clus schemes perform
similarly, the \clus scheme achieves its performance by selecting more appropriate \sls (as is evident in Fig.~\ref{fig:single_sl_latency}); this is not surprising, as
we limit the candidate super-links in the \naive scheme.

Fig.~\ref{fig:single_max} plots the maximum (rather than average) \eps generation 
latency over an $(s,d)$ pair. Here, we observe that the algorithms are not able to 
sufficiently decrease the latency of all the \eps requests; this is expected as the 
optimization objective of the algorithms is quite different. Nevertheless, we observe
that our main algorithm \geng still continues to outperform the other algorithms, 
and in particular, is able to lower the generation latency compared to the non-superlink scheme \nosl. 

Lastly, in Fig.~\ref{fig:single_sl_latency}, we plot the aggregate latency of the \epss generated over the SLs selected by various algorithms; in some sense, this metric represents how efficiently each algorithm is able to utilize the cost budget. We observe that while \naive and \clus results indicate under-utilization of the cost budgets, the \geng algorithm is able to utilize the available network resources more effectively (taking into consideration, its high performance in the previous two plots). We stipulate that this high-utilization is
due to the fact that we allow \geng to also select non-shortest paths between node pairs. 
%%%%%%%%

\softpara{Variants of Greedy Algorithm.}
We now consider two variants of our main Greedy algorithm \geng. 
See Fig.~\ref{fig:signle_greedy}.
First, we observe that {\tt GG-SP}, which is same as \geng except that it only considers shortest-paths as associated paths for super-links, performs similarly than our \geng algorithm when the cost budget is not high enough ($\leq30K$), but the performance gaps suddenly increases with higher cost budget. Intuitively, shortest paths are preferred for super-links as they yield the same reduction in generation latency of $(s,d)$ pairs with lower cost. But, with increased cost budget,
as more and longer super-links get selected, the intersections between candidate super-links increase and selection of non-shortest paths become important. 
%%%%%%%%%%%%%%%%%%%%%%%%%%%%%%%%%%%%%%%%
We also consider another variants of the greedy algorithm, viz., 
{\tt Pure-Greedy}, which is a pure-greedy algorithm in the sense that it does not delete any already-picked super-link; thus, at each stage, it only considers super-links that are disjoint with the already picked super-links.
% In addition, 
% {\tt Greedy-NDB} also uses just $\psi()$ as the metric for selection of a 
% candidate rather than the usual $\psi()$ per unit cost.
%%%%%%%%%%%%%%%
We observe that {\tt Pure-Greedy} significantly under-performs our main algorithm \geng; this observation validates the strategy of using both {\em Update} and {\em Append} options within \geng. 
%Finally, we observe that {\tt Greedy-NDL} outperforms the {\tt Greedy-NDB} scheme which validates our choice of selection metric. 

% \para{Multiple-Paths, Multiple-SLs Problem Results.}
% We now present performance comparison of various schemes for the Multiple-Paths problem. Here, we compare
% the following schemes: Iter-Greedy, and Iter-Heuristic. We still use one SL per path but with an additional constraint that an SL can only be used once per an $(s, d)$ pair.
% %%%%%%%%%%%%%%%%%%%%%
% See Fig.~\ref{fig:multi_latency}. As you can see, latency can be further reduced (compared to \sls) by using multi paths, multi SLs for establishing an \ep between two nodes. While multiple-paths itself reduce the latency, it helps the algorithms to pick better SLs (in terms of cost) to achieve to the requested latency.
% Iter-Heuristic performs better, in this case, as it has the advantage to access to all the paths for all the $(s, d)$ pairs in advance. This is the case while the SLs' cost (and even their complexity) is better in most of the cases. See Fig.~\ref{fig:multi_cost}-\ref{fig:multi_sl_latency}

             % 2 pages.
\section{\bf Conclusions and Future Directions}
\label{sec:conc}

In this paper, we proposed the approach of pre-distributing \epss to lower generation latency of expected \eps requests, and addressed the key optimization problem in this context. Our future work is focused on investigated more sophisticated variants and settings of the \sls problem, e.g., combining super-links and multiple-paths approaches, allowing multiple super-link in each entanglement path.
%to further enhance the high-fidelity entanglement rates.

\newpage
\bibliographystyle{IEEEtran}
\bibliography{main}
\newpage \newpage
\end{document}